# Evolution through time of Pyrrolizidine Alkaloids detection and quantification

Filipe Monteiro-Silva,[a] Gerardo González-Aguilar[b]

[a]Departamento de Química e Bioquímica, Faculdade de Ciências, Universidade do Porto, Rua do Campo Alegre, 4169-007 Porto, Portugal
[b]UOSE- INESC TEC, Departamento de Física, Faculdade de Ciências, Universidade do Porto, Rua do Campo Alegre, 4169-007 Porto, Portugal

______________________________________________________________________

*Abstract*

Pyrrolizidine Alkaloids (PAs) are a group of naturally occurring alkaloids that are produced by plants as a defense mechanism against insect herbivores. The analytical methodologies employed for their detection have come a long way since the first analytical experiment and in the last 30 years had an enormous development, both technological and experimental. It is notorious that before the generalization of certain technologies, especially in a post-war atmosphere, most scientific researches relied on what it is today thin-layer chromatography. Nevertheless this technique was not sufficient for accurately measure quantities and unambiguously identify compounds, therefore spectroscopic techniques arose as well as chromatographic techniques. While the first never really coped with PAs analysis requirements the latter, either as gas or liquid chromatography allowed the analysis of complex sample matrixes. Simultaneously, nuclear magnetic resonance also suffered significant developments while mass spectrometry has become an attractive technique due to increasingly higher maximum resolutions.

The observed tendency in recent years, in pyrrolizidine detection and quantification – as well as in many other areas – is that hyphenated techniques are the chosen methods. A large number of papers report multi-hyphenated methodologies, and the overwhelming majority relies on gas or liquid chromatography.

## Contents

## *1. Introduction*

Pyrrolizidines alkaloids (PAs) are a class of naturally occurring chemicals that are toxins biosynthesized by some plant species. More than 660 PAs have been identified from over 6000 plant species [1] that correspond approximately to 3% of the world's flowering plants and represent a convergent trait in the plant kingdom [2]. Species like the *Fabaceae* (Crotalaria), *Boraginaceae*, *Apocynaceae* (Echiteae) and *Asteraceae* (Senecioneae, Eupatorieae) occur in almost every habitat. They can be shrubs or vines, annual or perennials, and some species are invasive and considered as noxious weeds [3].

In terms of structure pyrrolizidines are two-fused 5-membered rings with a nitrogen atom at the bridgehead. An amino alcohol - necine - is the base system and variations on this core with typically highly branched and substituted acids originate mono, di or macrocyclic diesters of the unsaturated necine or otonecine bases [4]. Single esters at $C_9$, diesters at both $C_7$ and $C_9$ positions of the necine base can occur, sometimes with long cyclic diesters linking $C_7$ to $C_9$. The respective *N*-oxides (PANOs) can also be formed (Fig. 1) [5]. Many PAs frequently co-occur as the *N*-oxide and as the tertiary base. In plants an enzymatic catalyzed reaction originates *N*-oxidation of PAs, while a spontaneous reduction of PANOs occurs in the presence of biological or chemical reducing agents. There are records of the presence of tertiary PAs and PANOs even after decades of storage (under controlled light and humidity exposure), in lyophilized plant material [6].

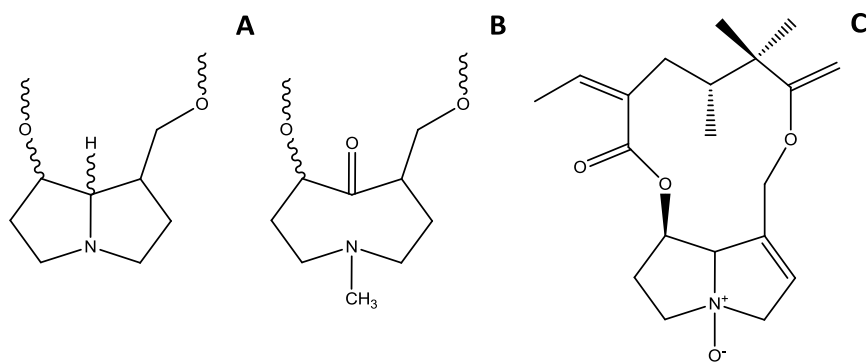

Figure 1. Pyrrolizidines structures. A - Necine base, B - Otonecine base and C - Senecionine *N*-oxide.

In general, PAs are optically active as they are chiral molecules and unsaturated PAs UV (ultra-violet) spectra usually show an absorption maximum *ca.* 214 nm [7]. At pH-values above 9 their esters can be hydrolyzed in aqueous solutions and the quaternisation of tertiary PAs can be promoted by halogenated solvents to the corresponding salts [8].

Property-wise, tertiary PAs are soluble in polar organic solvents, as well as in more lipophilic solvents like dichloromethane being slightly soluble in water while their protonation at the nitrogen atom can occur at low pH-values. On the other hand, PANOs are charged molecules that are partially soluble in water and in polar organic solvents.

Despite their specific stability, PAs and PANOs are prone to suffer influence from outside factors, resulting in changes on their structure, concentration or stability that could lead to changes to the PAs/PANOs ratio and/or total PA content of a determined sample. This is one of the reasons why caution should be taken when performing PAs and PANOs analysis. The main factor that may influence the results obtained is the temperature, as Hösch *et al.* reported in 1996. The time to which the samples were exposed to temperature severely influenced the PAs/PANOs ratio [9]. However, the temperature *per se* does not seem to influence these values but instead promotes the reductive activity of certain compounds over the *N*-oxides [10].

*1.1. Biological importance of PAs*

The commonly accepted definition of PAs also includes saturated pyrrolizidines even though their biological roles have not been thoroughly studied and are likely to be different [3]. Until date, PAs are classified as toxics and some are even classified as possibly carcinogenic to humans (Fig. 2) [11, 12]. This toxicity is associated to their metabolization. PAs can be converted to PANOs and their toxic form during digestion [8] and are themselves susceptible to be considered toxic [4].

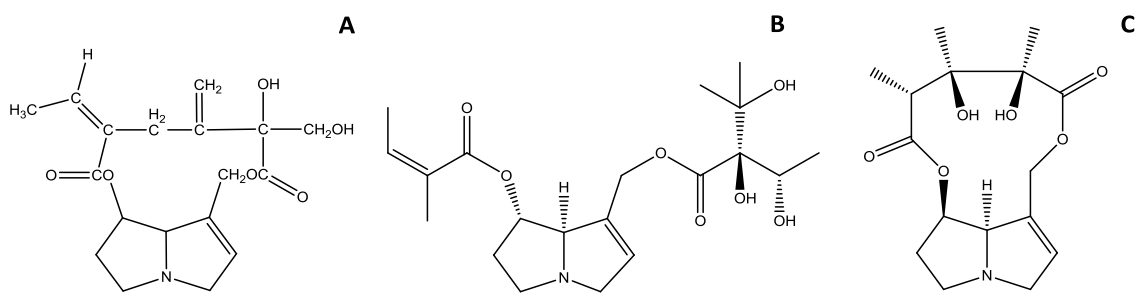

Figure 2. Examples of carcinogenic PAs: A - Riddelliine, B - Lasiocarpine and C - Monocrotaline.

The problematic of the PAs relies on the premise that exposure to this contaminant is not controlled and there is not a deep knowledge of their hepatotoxicity. Many of the PAs feature the necessary chemical properties to be hepatotoxic: a $C_1$-$C_2$ double bond on the pyrrolizidine moiety and a hydroxyl group able to undergo sterification [13].

As these xenobiotics enter the system, they are absorbed by the blood stream and enter the hepatic portal system and undergo metabolization. This metabolization has the purpose of converting a certain molecule, capable of crossing membranes, in one that can be excreted (through urine, sweat, etc.). Throughout the metabolic steps, the molecules' lipophilicity usually decreases, increasing their polarity. This step determines if the molecule proceeds to renal excretion or if it undergoes further metabolization [14].

If metabolization continues, they are due to be activated in the liver and in the case of PAs, to reveal their toxic effects. In the liver, esters and amides are susceptible to esterases action while nitro, azo and carbonyl moieties undergo action of reductases. By oxidase action, PAs transform into pyrrolic dehydro-alkaloids (dehydroPAs) which are reactive alkylating agents. These are believed to be responsible for liver cell necrosis and for hepatic sinusoidal obstruction syndrome [15]. Besides, there have been reports of additional damages such as lung vascular lesions characteristic of primary pulmonary hypertension [16]. DNA-binding capabilities have also been pointed as responsible for the genotoxic effects that pyrrolic metabolites possess (Fig. 3) [17, 18, 19].

Some PAs may present some degree of resistance to esterase due to the steric hindrance in the acid moiety. High levels of pyrrolic metabolites may be formed from the chain branch near the carbonyl groups, in the necine base. This transformation slows down possible hydrolytic steps. In the same manner, a spatial conformation of the basic moiety, which brings the ester groups closer, leads to mutual steric hindrance and subsequent hydrolysis inhibition of the resulting molecule [20].

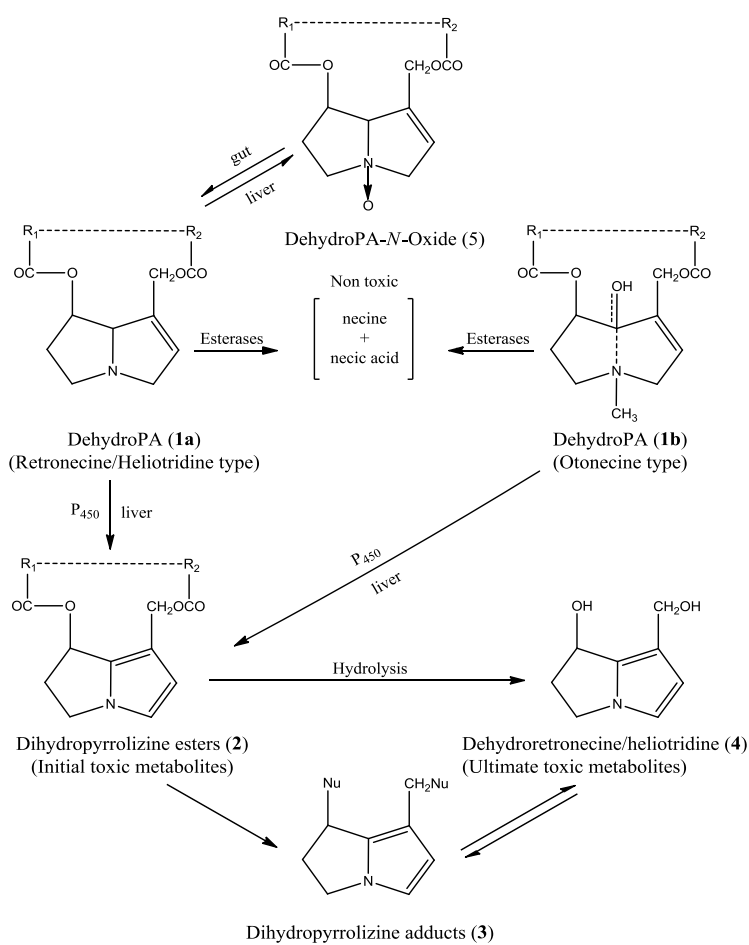

Figure 3. Summary of the metabolic pathway of dehydroPA (1a, b) by bioactivation and detoxification *in vivo*. Compounds (1a and b) are converted in hepatocytes, by Cytochrome $P_{450}$ (CYP) enzymes, to 6,7-dihydropyrrolizine esters (2). A fast nucleophilic (Nu) reaction occurs and adducts are formed (3), some of these adducts release dehydroretronecine/heliotridine (4) and the formation of adducts proceeds and may lead to chronic diseases. Compounds (4) can also be produced by a water reaction with initial metabolites (2). The hydrolysis of (1a, b) ester groups can occur by detoxification through esterases and non-toxic necid acids are produced. In the liver *N*-oxides (5) are obtained by *N*-oxidation of (1a) and are reduced in the gut to the free base form. R1-R2 are a subgroup of dehydroPA diesters [21].

The rate to which pyrrolic metabolites are formed is influenced by the induction or inhibition of the mixed-function oxidases in the liver, but evidences of a relationship between the rate of metabolism and expression of toxicity are still not known [21].

Despite the limited information on the dose-response relationships, according to the World Health Organization, intake of PAs may present a health risk and exposure should be minimized whenever possible [16].

## *2. Analytical methods for pyrrolizidine identification and quantification*

### *2.1. Sample preparation*

Prior to determine any sample content, one must consider the variety of PAs that exist and the possibility of co-occurrence of PANOs also. Besides simultaneous presence, both matrix and matrix interferences have always to be taken into account.

In order to obtain an efficient extraction of both types certain characteristics should be acknowledged:

- PAs and PANOs are polar compounds;
- PAs bearing basic nitrogen are suitable to be extracted via classic alkaloid methodology with acidified aqueous conditions or polar to semi-polar organic solvents;
- PANOs are oxidized forms of parent PAs and, if considered adequate, can be reduced to them. The most common method is a reaction with elemental zinc in acidic aqueous solution with subsequent extraction [22]. Otherwise they can be extracted by polar solvents or diluted aqueous acids;
- A cleanup step, either before or after the extraction can be performed using, for example, non-polar solvents to suppress the interferences of any non-polar compounds.

The type of extraction procedure needs to be assessed depending on the matrix available. Solid samples must have a different treatment regarding a liquid sample. If it is sought to extract PAs/PANOs from tissues or heterogeneous samples, homogenization should be assured prior to extraction in order to obtain a representative sample. Isolating PAs from animal tissues might be difficult as they might be metabolized very rapidly once ingested. Yet, several types of samples were successfully analyzed up-to-date: honey [23], milk [24], eggs [25], urine [26] and blood [15] are just a few examples.

The extraction process is also dependent of the type of analytical technique(s) chosen, adopting a more iterative stance. It is known that PANOs are thermophobic and become unstable at high temperatures – e.g. those needed on GC analysis – therefore direct analysis of samples containing PANOs must not be done without previous derivatization [27]. Soxhlet extraction is one of other commonly used techniques [9, 28, 29], requiring high temperatures and boiling solvent(s), which allows to obtain high recover rates, but if extraction of PANOs is required then this technique should be avoided because of the motives aforementioned.

Another commonly adopted technique on sample preparation and extraction is the use of solid-phase extraction (SPE). Several types of materials are available for usage in PAs/PANOs extraction from different sample types: from typical $C_{18}$ [30], Ergosil [31], LiChrolut [32], Serdoxit [33], Extrelut [34] or SCX (strong cation-exchange) [35].

*2.2. Qualitative analytical methods*

Several techniques have been used and developed over the years in this field and will be discussed below, through a chronological timeline of the last 30+ years.

Nevertheless, a brief reference goes to the first technique that sought to identify PAs and PANOs already in the 1960's by distinguishing their different polarities and oxidation properties: paper chromatography. By using a rudimentary version of modern thin-layer chromatography, Arthur Thurlby Dann used acetic anhydride in order to obtain selective and colorful spots [36]. A few years later Robin Mattocks enhanced the previous method, using hydrogen peroxide or peroxide anhydride reagents, making it sensitive for alkaloids with an unsaturated pyrrolizidine ring [37].

## *3. Quantitative methods*

It was only after 1980 that pyrrolizidine alkaloids began to be quantified and for that purpose, several techniques were used. In the following pages a review of the most important reports and progresses of the quantitative methods is presented and the study will be divided in three main periods. These periods are 1980-99, 2000-09 and finally 2010-13, as these time intervals represent - for us - different development stages in both analytical techniques and pyrrolizidine alkaloids analysis.

*3.1. From 1980 to 1999*

*3.1.1. High Performance Liquid Chromatography and Liquid Chromatography-Mass Spectrometry*

High Performance Liquid Chromatography (HPLC) is a resourceful and widely used analytical technique that being non-destructive is useful when limited amounts of sample are available and their recovery is required. In the case of PAs/PANOs, simultaneous determination is possible without any prior derivatization steps but limited and unspecific information is obtained [7].

In the referred time span, HPLC was not the preferred method of choice when PAs analysis was needed, at least not in the first few years. Few papers with HPLC determination were published probably because, among other reasons, to the lack of a chromophore that enabled their detection with conventional ultra-violet systems. Adding such a group through derivatization, either before or after the chromatographic column, would allow HPLC detection but would also add complex steps to the process of analysis which desirably, should be fast and simple. Also the generally low solubility in organic solvents that do not portray hydroxyl groups disabled the development of adequate solvent system–column combinations [27].

In 1981, Ramsdell and Buhler reported HPLC analysis of *Senecio vulgaris* and *S. jacobaea* flower tops for PAs [38] using a standard $C_8$ reverse-phase (RP) column. A few years later, in 1986 Kersierski and Buhler improved a HPLC method using a styrene-divinylbenzene column and ultraviolet detection allowing the simultaneous determination of the senecionine, senecipylline, and retrorsine as well as their major metabolites produced during *in vitro* transformation of PAs [39]. With the evolution of column packing in the decade of 1990 some very interesting works were done like the one of Brown *et al.* [40] using cyano and phenyl-bonded columns allowing shorter running times. In 1997, Crews and his co-workers developed a method that combined SPE with HPLC-MS detection for PA determination in honey samples of Ragwort (*Senecio jacobaea*). The advantages of the method included the determination of individual alkaloids and a considerable improvement in specificity, sensitivity as well as speed-wise. The lowest detection limit achieved in honey samples was 0.002 ppm [34].

*3.1.2. Gas Chromatography and Gas Chromatography-Mass Spectrometry*

Gas Chromatography (GC) allows the use of samples, either in the form of extracts or separate components, however in the first case it is necessary a preliminary step of ion-exchange and for some samples a previous modification or derivatization is also required. Analysis of PAs samples is possible without any prior modification however for the analysis of PANOs, due to their properties (high polarity, low volatilization ability and instability at the temperatures required to GC analysis), they should be converted to the corresponding tertiary PAs or derivatized. The derivatization process allows PANOs to become less polar, which can be achieved by introduction of:

- bulky, non-polar silyl groups to form trimethylsilyl ethers [41] or

- boronate reagents, like methyl, butyl or phenyl boronic acids, to form bonds across vicinal diol groups [42], [43].

The apparent disadvantage of derivatization is that it may cause some degree of destruction or modification to senecionine and seneciphylline [44] however it allows the determination of retronecine base [45].

It was not until 1982 that GC-MS was used in the determination of PANOs. In the work developed by Brauchli *et al.* 7-pyrrolizidine alkaloid-*N*-oxides were determined by this methodology [28]. A few years later, in 1988 Hendriks *et al.* used positive and negative ion-chemical ionization to detect PAs in *Anchusa officinalis* after trimethylsilylation [46]. Pieters and his co-workers opted to quantify PAs from *Senecio vernalis* using GC and NMR ($^1$H and $^{13}$C) after reducing PANOs to the parent tertiary bases [47]. The first report of PAs determination from *Senecio serra*, *S. dimophophyllus* and *S. hydrophyllus* was published by Stelljes *et al.* in 1991, and it allowed through GC-MS analysis to profile and compare the *Senecio* species in terms of PA content [48]. The method employed a medium-low polarity DB-17 column that afforded a less time-consuming identification of PAs in comparison with NMR analysis. A very interesting work was performed by Dueker *et al.* where PAs where incubated in guinea pig carboxylesterase (GPH1), with the purpose of measuring the release of retronecine (RET). The GC analysis provided information on the release of RET by enzymatic hydrolysis or base-catalysis processes that occurred on the parent molecules [49]. Hovermale and Craig established a routine method for RET determination using GC after derivatization with bis-(heptafluorobutyrate). With this work it became clear that GC started to provide very remarkable results in terms of limit detection [45].

*3.1.3. Enzyme-Linked ImmunoSorbent Assay (ELISA)*

Enzyme-Linked ImmunoSorbent Assay had its first steps *ca* 1960 as radioimmunoassays performed by Yalow and Berson [50] but the radioactivity was a downside of the technique. Since enzymes react specifically under certain conditions, this property was explored - under the advent of solid-phase organic synthesis by Robert Bruce Merrifield [51] - binding enzymes to antigens or antibodies that were fixed to a solid surface. One of the first reports appeared in 1966 [52]. In general terms ELISA involves the production of visible color shifts associated to certain chromogenic reporters and substrates that stress the presence of analytes or antigens.

Few reports during this time period arised comprising ELISA assays on PAs. The first one, in 1989, by Bober and her team demonstrated how using the common structural necine base, it was possible to raise antibodies, through retronecine-protein (bovine serum albumin - BSA), to detect retronecine as well as retrorsine, senecionine, and seneciphylliine in a competitive inhibition enzyme-linked immunosorbent assay [53, 54]. Later, the same group reported the production of retrocine-moiety class-specific antibodies, this time through retronamine-BSA conjugate [55]. In 1992, David Roseman and his group achieved very interesting results developing a class-specific method claiming detection limits of 1.0-100 ppb, comparable at the time to GC/GC-MS detection limits [56]. In 1996, the same group achieved detection limits of 0.5-10 ppb for retrorsine [57]. All these papers reported the existence, to some degree, of mixed crossed-reactivity as they tend to suffer influence from other PAs and/or PANOs (see section 3.2.3).

*3.1.4. Other techniques*

Other techniques that allowed scientist to research PAs in this time span were mainly focused on other types of chromatography.

In 1980, curiously published in the same issue of the same journal, both Huizing and Molyneux claimed findings on a method to determine the presence of PA on TLC plates (thin-layer chromatography) using chloranil [58, 59]. Other methods included oxidation either by Dragendorff's or Erlich's reagent sprayed on TLC plates [60] or ion-pair adsorption chromatographic separation developed by Huizing and Malingré [61].

On the same year, Birecka *et al.* published a very simple and fast method that through the stoichiometric reaction of protonated alkaloids with methyl orange permitted the spectrophotometric assessment of alkaloids up to a concentration of 0.5 ppm [62]. A novel method for extraction of PA was achieved by Schaeffer and his collaborators employing supercritical fluid in 1989 [63] and two years later Bicchi *et al.* tested the applicability of offline supercritical fluid extraction of PAs for identification with GC, with higher recoveries in comparison to typical Soxhlet extractions [64]. By 1992, Roeder and colleagues improved an existing method to, by spectrophotometric means, reach detection limits up to 10 ppm in 1 gram of substrate [65].

As a final example, counter-current chromatography (CCC) – not a very common separation technique - was employed by Huxtable and co-workers in 1996 to separate and purify PAs from *Amsinckia tessellata*, *Symphytum spp.*, *Trichodesma incanum* and *Senecio douglasii longilobus* [66]. CCC consisted in one liquid to act like a stationary phase - sustained against the outer walls of a helical column due to its rotation – and a mobile phase to be pumped through the system. This method though it presented low resolution and separation ability to today's standards (350 to 1000 theorical plates), had the advantage of minimal contamination and adsorptive loss due to the absence of a solid support.

*3.2. From 2000 to 2009*

*3.2.1. High Performance Liquid Chromatography and Liquid Chromatography-Mass Spectrometry*

After the year 2000, liquid chromatography and related techniques applied to the determination of PAs or PANOs grew rapidly, mostly associated to the developments and breakthroughs on column packing and processing. The use of mass spectrometry (MS) coupled to HPLC became more common and an unequivocal method for identification.

In the study of 2001 by Chou *et al.*, the development of a sensitive and reliable $^{32}$P-postlabeling and HPLC method for detection of (i) two DHR-3'-dGMP and four DHR-3'-dAMP adducts and (ii) a set of eight DHR-derived DNA adducts *in vitro* and *in vivo* was achieved [67]. A simultaneous study performed by the same group suggested that riddelliine (Fig. 4) might induce liver tumors in rats through a genotoxic mechanism and the DHR-drived DNA adducts might contribute to the development of liver tumor [68].

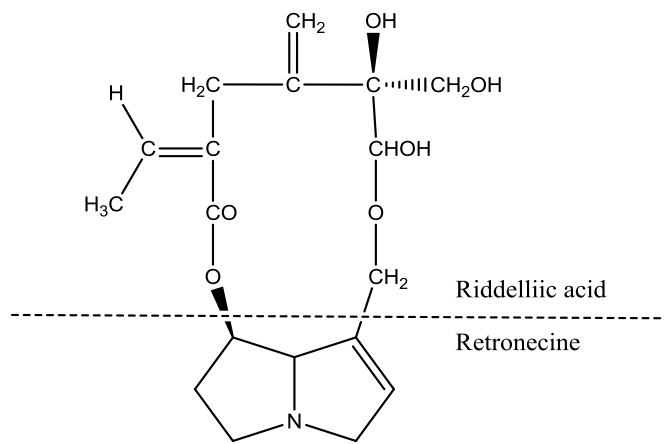

Figure 4. Chemical structure of Riddelliine.

In the same year, Glowniak and collaborators applied a cation-exchange solid-phase extraction to the process of PA extraction with very promising results. They were able to simultaneously separate both free bases and *N*-oxides with recovery rates of 80-100%, followed by gradient ion-pair HPLC (reverse-phase $C_8$ 5-μm column) [69]. A few years later they also reported [70] a new procedure to separate/identify PAs in various plant extracts which involved a rapid resolution $C_{18}$ column and MS fragmentation and again, in 2004, the structure characterization of *Onosma stellulatum* and *Emilia coccinea* with a RP HPLC with ion-trap MS [71], and in 2006 *Symphytum cordatum* alkaloids [72]. Colegate *et al.* reported honey samples analysis by SPE followed HPLC identification [73] and quantification through internal standard normalization while Bligh and colleagues preferred to derivatize existing PAs, evaluating a common retrocine marker (7-ethoxy-1-ethoxylmethyl retronecine derivative), produced by the different retronecine esters-type pyrrolizidine alkaloids (RET-PAs) [74]. Two years later, a slightly improved method was reported by Wang *et al.* in which derivatization and elution conditions were optimized [75]. A different extraction process was presented by Ong *et al.*, combining microwaves and hot water after which identification was performed using mild elution conditions. Further identification resorted to LC-MS, using a 3-µm $C_{18}$ column and good relative standard deviation (RSD) values of precision were obtained [76]. Later in the same year an improvement of the same method was attempted using sonication instead of microwave-assisted synthesis. Pressurized hot water extraction (PHWE) was also investigated but lycopsamine extraction efficiency was considerably reduced with the increase of temperature (Fig. 5). According to the authors, the lower efficiency might have been due the presence of dissolved nitrogen with effects on lycopsamine solubility or leading to its oxidation, resulting in high RSD values [77].

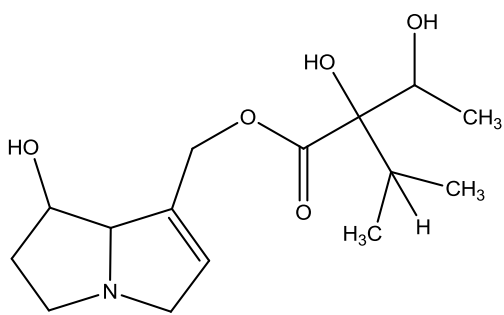

Figure 5. Chemical structure of Lycopsamine.

The analysis of PAs in *Jacobaea vulgaris* was also investigated by Joosten *et al.* in 2009. In a very insightful work, their objective was to determine the applicability of different extraction and detection techniques, namely GC with nitrogen phosphorus detection (GC-NPD) and LC-MS-MS. The detection of PAs by both techniques was comparable - despite the apparent 15 fold difference in detection limit (LOD) - but LC-MS outperformed GC due to the unnecessary reductive step for GC and allowed the detection of 11 additional different PANOs. This fact might indicate that regardless the reductive strategy, some information might be lost during that process [78].

*3.2.2. Gas Chromatography and Gas Chromatography-Mass Spectrometry*

Gas chromatography also benefited from technological evolutions in this time period, not only on column efficiency and sensitivity but also on tandem analysis techniques, namely mass spectrometry. Their sensitivity and LOD as well as their dynamic range were gradually enhanced as ionization techniques became finer and detectors gained better resolution.

A very exciting work, in 2000 by Schoch, Gardner and Stegelmeier employed GC-MS-MS for PA metabolite detection, using a protein-metabolite conjugate as a pseudo-standard for the calibration curve [79]. They studied riddelliine supplementation effects on pigs - generally accepted as the most suitable human mimetic model – by analyzing both liver and blood pyrrolic metabolites. Differences in metabolite levels could be discerned, but amounts of riddelliine fed and detected did not show correlation. In 2003, Karlberg and Wretensjö assessed PA content in borage oil and the effect of the refinement process, with a combination of two different columns for GC-MS analysis: a non-polar DB-1 and a mid-polarity DB17; and a 100% dimethylpolysiloxane column for GC-FID [80]. The authors did not find presence of any PAs except the crotaline they added as a reference and concluded that no PAs were present at a level above 100 ppb. In 2008, two interesting papers were published. The first, by Schreier *et al.* regarding honey sampling analysis through GC-MS with previous SPE and derivatization in which the authors claim a limit of quantification (LOQ) of 0.01 ppm [81] and the second by Hartmann *et al.* where assignment of stereoisomeric 1,2-saturated necine bases was achieved. In the latter, two different equipment configurations were used, one with a low polarity column and the second with a slightly more polar one: 100% dimethylpolysiloxane versus 95% dimethylpolysiloxane/5% phenyl. The authors announced unambiguous identification achieved by detailed GC–MS analysis and confirmation of the structures by NMR [82].

*3.2.3. Enzyme-Linked ImmunoSorbent Assay (ELISA)*

Recently, ELISA technologies evolved to use fluorescent probes, in addition to the previously used chromogenic markers as well as electrochemiluminescent ones which brought higher sensitivities and higher sample throughput [83, 84].

Lee and his collaborators investigated [85] crossed reactivity among 16 pyrrolizidine alkaloids and found that this interaction between the *N*-oxide and the free base forms allowed an estimation of the total PA content in the sample and inferred that the methylene group at the $C_{19}$ carbon was the primary antigenic site and promoter of the immunologic response for the antibodies raised against the used immunogen.

*3.2.4. Other techniques*

Despite the main identification and quantification techniques used were gas and liquid chromatography, mostly with mass-spectrometry support, some other techniques continued to be evaluated to assess pyrrolizidine alkaloids and their *N*-oxidized bases.

In 2004, a group of Brazilian researchers led an investigation to determine the cause of livestock deaths in several outbreaks [86]. Upon PA poisoning suspicions, samples were collected and after liquid extraction and oxidation by Erlich's reagent, identification was performed by TLC and quantification by UV spectrophotometry. Coincidentally at the same time, Khan, Molyneux and Schaneberg developed a reverse-phase HLPC with evaporative light scattering detection (ELSD) which, according to the authors, should be able to detect simultaneously PAs with and without a chromophore. A rather simple elution gradient was set by the authors, at room temperature that enabled a LOD of 40 ppm [87]. Dickinson *et al.* in 2005, hydrolyzed PAs to its parent base retronecine and proceeded to fluorination obtaining four derivatives after which they were injected on MS [88]. As a result of the insertion of electron withdrawing fluorinated groups, the preferred fragmentation changed from the nitrogen α-cleavage to a charge site migration, resulting in an alkyl–oxygen bond cleavage and the formation of a stabilized allylic cation. In a slightly different investigative area, Pothier and Galand applied automated multiple development in thin-layer chromatography to opiate alkaloids which, not sharing the same properties, might have a similar potential of applicability [89]. Barnett, Gorman and Bos in 2005 published an outstanding work combining chemiluminescent reactions with PAs, using ruthenium's remarkable properties (Fig. 6) [90]. Performing flow-injection and sequential-injection analysis (FIA and SIA respectively), the authors were able to determine PAs presence with good LOD. Another potentially useful technique for PAs analysis was investigated by Li *et al.* in 2005 based on micellar electrokinetic chromatography (MEKC) [91]. The authors claimed that, in a single 17 minute run they were able to separate a senkirkine, senecionine, retrorsine, and seneciphylline containing matrix. The presented LOD values of 1.19 to 2.70 ppm were not as good as in others techniques (e.g. chromatographic techniques) but nevertheless it was a simple and rapid method that provided good results. Simultaneously Li and Yu published another study on PA detection that consisted on dynamic pH junction-sweeping capillary electrophoresis (CE) [92]. The hyphenization of CE with dynamic pH junction-sweeping allowed the authors to enhance the CE detection sensitivity to values as low as 30 ppb, which enabled them to reach good results.

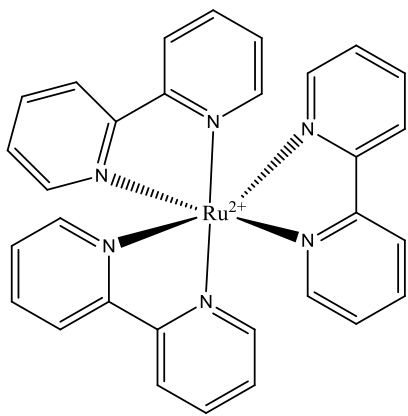

Figure 6. Structure of complex ion tris(2,2'-bipyridyl)ruthenium(II).

*3.3. From 2010 to 2013*

*3.3.1. High Performance Liquid Chromatography and Liquid Chromatography-Mass Spectrometry*

In recent years, liquid-chromatography has seen quite a few technical developments. However, the emergence of Ultra-Performance Liquid Chromatography (UPLC) might be considered a milestone in chromatographic technology. Along with and because of UPLC, new types of columns were developed with particle sizes sub-2 µm, namely 1.7 and 1.5 µm. These new columns allow solvent pressures up to 1000 bar or more and an increase in the resolutive capabilities with valuable run time decrease.

Xu *et al.* have developed an UPLC system coupled with tandem mass-spectrometry (MS-MS) on a tandem quadrupole mass-spectrometer [93]. They have incorporated a precursor ion scan (PIS) acquisition system with multiple reactions monitoring (MRM). Successfully they were able to detect pairs of characteristic product ions at m/z 120/138 or 168/150, specific to retrocine and otonecine type PAs (Fig. 7).

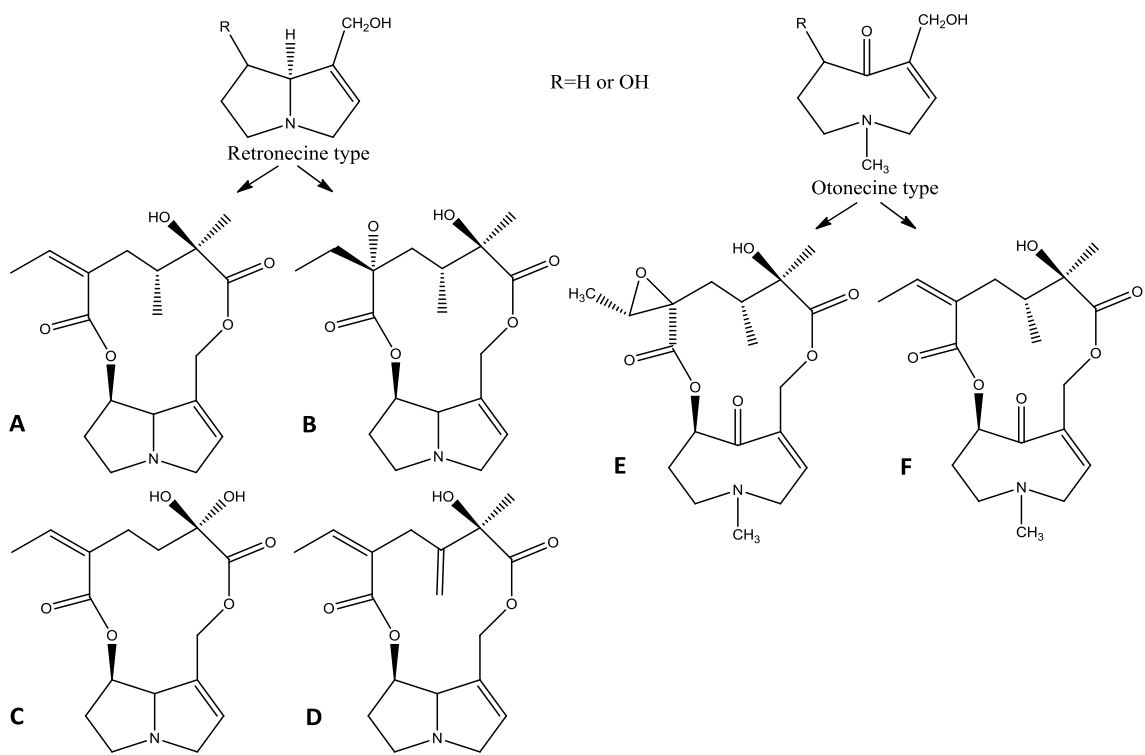

Figure 7. Retronecine-type and otonecine-type PA toxins representative chemical structures used in the study [93].
A: Senecionine, B: Jacobine, C: Retrorsine, D: Seneciphylline; E: Otosenine and F: Senkirkine.

Following a previous study [17], referred in section 3.2.1., Fu *et al.* reported the development of specific HPLC-MS-MS method that enabled DHP-derived DNA adducts detection [19]. The method allowed the quantification of DHP-2′-deoxyguanosine (dG) and DHP-2′-deoxyadenosine (dA) adducts level, by MRM analysis with isotopically labeled internal standards (DHP-dG and DHP-dA). This method granted the authors a further insight on adduct formation allowing them to propose a general metabolic activation mechanism. A quite extensive work was done by Dübecke, Beckh and Lüllmann in 2011, analyzing over 4000 samples searching for PA contamination [94]. After SPE, identification was performed via LC with tandem MS which unveiled PAs in 66% of the bulk, unpacked honey and in 94% of the retail honey analyzed. Such bulk amounts of sample analysis were achieved using a UPLC-comparable column (under 2 μm size particles), which allowed very short run times of 10 minutes and LOQ varying from 1 to 3 ppb, as claimed by the authors. Quite an interesting work was performed by Orantes-Bermejo *et al.* in the same year, using HPLC-MS with electrospray ionization to determine total pollen content percentage of *Echium spp.* pollen as well as simultaneous measurements of PAs and PANOs [94]. While PAs identification was performed by typical HPLC procedures, PANOs identification was executed by MS analysis, namely [M + H]$^+$ adduct ion and dimer adduct [2M + H]$^+$ formation. Another quite extensive work was the investigation performed by Mol *et al.*, in which PAs presence (among other substances) was investigated [95]. Using a LC-MS system possessing a mass resolving power of 100,000, a high mass accuracy was to be expected hence non-selective sample preparation and non-targeted data acquisition was performed. Despite being a quite self-critical paper, LOD were achieved in the range of 0.01 to 0.05 ppm. In 2012, a very good comparative study with two different liquid chromatography systems was performed by Khan and colleagues, with the purpose of simultaneous determination of sesquiterpenes and pyrrolizidine alkaloids from rhizomes and dietary supplements [96]. An UPLC with UV detection and a HPLC with time-of-flight MS methods were successfully developed with interesting results. They obtained LODs for PAs detection of 5 ppm for the UPLC-UV system and 0.1 ppm for the HPLC-TOF-MS. The differences between the methods employed in these experiences led to a major expressive result – the latter was 5000 times more sensitive than the former. The essence of that difference relied on the fact that the combination of retention times and exact masses might offer an unequivocal identification of the compounds. This work showed that while the UPLC-UV provided speed of analysis (with shorter equilibration times) with lower solvent consumptions and resolution, HPLC-TOF-MS provided high acquisition speeds and precise mass measurements as well as full scan spectral sensitivity. One year later, Letzel and Aydin applied a slightly different method for simultaneous determination of the PAs, PANOs, sesquiterpenes and their derivatives [97]. The authors combined two different ionization techniques - electrospray ionization and atmospheric pressure chemical ionization (APCI) - which provided ionization of molecules with different polarities, broadening the polarity range and permitting ESI unresponsive molecules to be ionized by APCI. Another bulk-sample detection method was developed by Furey *et al.*, which allowed toxic retronecine and otonecine-type PAs to be recognized by comparison to reference compounds via a spectral library, using liquid chromatography with ion trap mass spectrometry [98]. Values for LOD and LOQ varied from 0.0134 to 0.0305 and 0.0446 to 0.1018 ppm, respectively. Seraglia and colleagues reported a specific and rapid method to evaluate the presence of PAs in *Borago officinalis* seed oil [99]. Though no PAs were detected on the unspiked samples, spiking controls ahead of direct MS injection allowed this workgroup to affirm they had reached LOD in the order of 200 ppt.

*3.3.2. Gas Chromatography and Gas Chromatography-Mass Spectrometry*

Within this time frame the number of papers using gas chromatography instead of liquid chromatography has decreased. Contributing to this tendency might be the absence of unequivocal or commonly accepted methodology that enable detection and/or quantification of PAs simultaneously to PANOs, without loss or corruption of information. Derivatization, as stated before (see section 3.2.1), might induce impaired data, therefore and despite gas chromatography still being one extremely valid analytical technique, it is mostly used when coupled with mass spectrometry, becoming a more robust technique.

In 2011 Beuerle *et al.* published an interesting paper in which a comparison between two different analytical methodologies was made [100]. High-resolution GC-MS was performed with selective ion monitoring towards signals m/z 93, 183 and 299 while LC-MS was performed in ESI positive ion mode. Both methods had pre-injection treatment: in the first case zinc reduction, SPE, reduction to necine base and derivatization; in the latter, QuEChERS methodology was applied – used by Anastassiades *et al.* for the determination of pesticides in food matrices [101]. Later Beuerle and Cramer performed another very interesting work by analyzing antibacterial honey, used for wound care, e.g. burns, minor abrasions and surgical wound [102]. The authors have used GC-MS analysis to identify PAs content in the collected samples after reduction and derivatization procedures. Quantification was done by monitoring m/z 183 signal (corresponding to the base peak) and the values obtained were representative of the total amount of 1,2-unsaturated retronecine-type PAs per sample.

### 3.3.3. Other techniques

An NMR-based metabolomics study was performed by Leiss *et al.* in 2011. It comprised the study of metabolites of plants, using uni- and bi-dimensional NMR assays in order to resolve structures of possible compounds present in samples, simultaneously profiling the metabolome of a certain plant. The authors stated also that this type of studies might allow future resistance breeding and biopesticide development [103]. Another interesting, non-destructive method for PA analysis was developed by Carvalho *et al.*, based on near infra-red (NIR) analysis [104]. This type of study is very promising as it might pose as an alternative to other, more complex and elaborate methods, even though its state-of-the-art is not as accurate and sensitive.

## 4. Promising Techniques

*Molecularly Imprinted Polymers (MIPs: The future analytical method?)*

Other technologies might contribute to PA analysis and one that may have that potential is Molecularly Imprinted Polymers (MIPs). MIP is a fast-developing technique introduced around 1931 and it is based on lock-and-key configurations, although MIPs do not identify the exact structure of the analyte. The essence of the process is quite simple: starting from a template molecule on which a polymerization process is initiated and at the end it is removed, leaving a cavity on the polymer (Fig. 8). This cavity on the polymer has the configuration of the initial template molecule, which can be used to detect other molecules that match the same configuration (hence the lock-and-key depiction). MIPs can have partial positive response when interacting with molecules which are fragments of the original target and for this reason the results are not 100% unequivocal. Two main processes for obtaining MIPs are currently used: Wulff's covalent bonding [105] or Mosbach and Mayes' non-covalent bonding [106]. With Wulff's technique a template-monomer complex is covalently formed, polymerized and undergone selective breaking of said covalent bond. On Mosbach and Mayes' method a high degree of crosslinked polymer interacts with the template molecule, either by non-covalent bonds, hydrogen bonds or electrostatic interactions. The template is removed and molecular recognition site(s) is(are) created. Despite different methodologies, both techniques are able to produce polymers with functional recognition sites.

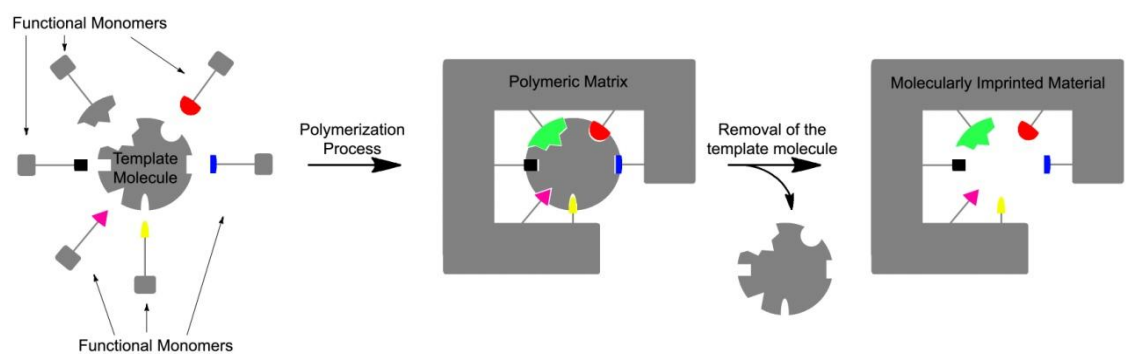

Figure 8. Summarized basic steps for Molecularly Imprinted Polymer synthesis.

Its potential has recently allowed several studies using fungi [107], viruses [108], proteins [109, 110], peptides [111, 112], nucleic acids [113] or even organic molecules [114]. Mosbach and Kempe already in 1995 [115] published a remarkable paper reporting the successful preparation of molecularly imprinted stationary phases for chromatography use on the separation of amino acids, peptides and proteins, while a few years later reported new enzyme inhibitors [116].

Therefore with MIPs technique it is possible to develop tailor-made polymers with selectiveness towards a desired compound or group of compounds, with potential applications in almost every area of scientific investigation and development.

## 5. Conclusions and final remarks

There are several techniques available nowadays that enable PAs analysis and therefore there is not an universally or widely accepted methodology.

Since PAs and PANOs are (metabolically) interconvertible, it is necessary that both species are included in the analytical determinations and there are still some limitations regarding that issue. There is also a lack of easily accessible standards for comparison or reference which inhibits the development of a methodology that could allow further insight on metabolic, genotoxic and toxicokinetic pathways. Even though there have been quite a few milestones in PAs detection and quantification, it is notorious that since around 1990, no scientific innovations have been done in this area, only the technology and analytical apparatuses did, which in turn allowed lower and lower detection and/or quantification limits.

From an analytical perspective, the clear trend is to use gas or liquid chromatography, mostly in mass spectrometry hyphenated systems to overcome existing handicaps and generate unequivocal results.